\documentclass[12pt]{iopart}
\usepackage[a4paper, total={6.5in, 10in}]{geometry}

\usepackage[english]{babel} 

\usepackage{microtype} 

\usepackage[svgnames]{xcolor} 

\usepackage{booktabs} 

\usepackage{lastpage} 

\usepackage{graphicx} 
\usepackage{subfigure}

\usepackage{enumitem} 
\setlist{noitemsep} 
\usepackage{xcolor}

\pdfoutput=1
\usepackage[T1]{fontenc}
\usepackage{float}
\usepackage{caption}
\usepackage[T1]{fontenc}

\usepackage{geometry}

\begin{document}

\title[L-shell X-ray conversion yields for laser-irradiated tin and silver foils]{L-shell X-ray conversion yields for laser-irradiated tin and silver foils}

\author{R. L. Singh\textsuperscript{1}, S. White\textsuperscript{1}, M. Charlwood\textsuperscript{2}, F. P. Keenan\textsuperscript{1}, C. Hyland\textsuperscript{2}, D. Bailie\textsuperscript{2}, T. Audet\textsuperscript{2}, G. Sarri\textsuperscript{2}, S.J. Rose\textsuperscript{3}, J. Morton\textsuperscript{4}, C. Baird\textsuperscript{5}, C. Spindloe\textsuperscript{5} and D. Riley\textsuperscript{2}}

\address{\textsuperscript{1}Astrophysics Research Centre, School of Mathematics and Physics, Queen's University Belfast, University Road, Belfast BT7 1NN, UK\\ 
	\textsuperscript{2}Centre for Plasma Physics, School of Mathematics and Physics, Queen's University Belfast, University Road, Belfast BT7 1NN, UK\\ 
	\textsuperscript{3}Blackett Laboratory, Imperial College, London SW7 2BZ, UK\\ 
	\textsuperscript{4}AWE, Aldermaston Reading RG7 4PR, UK\\ 
	\textsuperscript{5}Science and Technology Facilities Council, Rutherford Appleton Laboratory, Harwell Campus, Didcot, OX11 0QX \\}
	
\ead{raaj.phys@gmail.com}

\begin{abstract}
We have employed the VULCAN laser facility to generate a laser plasma X-ray source for use in photoionisation experiments. A nanosecond laser pulse with an intensity of order ${10}^{15}$ W{cm}$^{-2}$ was used to irradiate thin Ag or Sn foil targets coated onto a parylene substrate, and the L-shell emission in the $3.3-4.4$ keV range was recorded for both the laser-irradiated and non-irradiated sides. Both the experimental and simulation results show higher laser to X-ray conversion yields for Ag compared with Sn, with our simulations indicating yields approximately a factor of two higher than found in the experiments.  Although detailed angular data were not available experimentally, the simulations indicate that the emission is quite isotropic on the laser-irradiated side, but shows close to a cosine variation on the non-irradiated side of the target as seen experimentally in previous work.
\end{abstract}


\section{Introduction}
\label{sec:intro}

Over the past five decades, various types of X-ray sources have been produced by irradiating solid and gaseous targets with high power lasers in controlled laboratory settings. These X-ray sources are useful in research fields such as inertial-confinement fusion, plasma diagnostics (scattering, radiography, fast ignition and absorption spectroscopy) \cite{Glenzer2009, Woolsey1998,Bradley1987, Lindl2004}, laboratory astrophysics \cite{Foord2004, White2018} and warm dense matter. Many experiments have been performed to study multi-keV X-ray sources in which various elements have been investigated for K-shell \cite{Phillion1986, Fournier2004, Riley2002, Hu2007}, L-shell \cite{Back2003, Kettle2015} and M-shell \cite{Kania1992} emission. The X-ray emission duration, conversion efficiency of laser energy to X-ray energy, and the dependence of conversion efficiency on foil thickness have been investigated, with the latter also depending on the laser and target parameters in a given X-ray energy range \cite{BAILIE_2020, WANG2021, Hu2008}.  
Much of this previous research studied the X-ray emission coming from the laser-irradiated (front) side of the target foil. However, in many instances, the investigation of X-ray emission observed on the non-irradiated (rear) side of the target foil can be very useful, for example, when employing this X-ray emission to photoionise or volumetrically heat a gas or foil target \cite{White2018, Kettle2015}. 

In previous work, we investigated the conversion efficiency for L-shell emission for Sn foils \cite{BAILIE_2020} irradiated with 351 nm laser radiation. We studied the efficiency as a function of angle with respect to the target normal, and its dependence on the foil thickness. 

In this paper, we report measurements of L-shell emission for Ag foils \cite{Kemp2015} and compare these to results for Sn foils under the same irradiation conditions. The measurements were taken as a part of experiments on the VULCAN laser at the UK Central Laser Facility, sited at the Rutherford Appleton Laboratory in Oxfordshire. These experiments were to study the production of photoionised plasmas relevant to accretion-powered astrophysical X-ray sources. We used an Ag foil rather than Sn as we expected a higher L-shell flux, due to the lower Z of the former. Accurate knowledge of the X-ray emission from the foil is vital in understanding the dynamics of the photoionised plasma experiments. Hence we dedicated some experimental time to investigate the laser to X-ray conversion yield, and the results are presented here for the benefit of future researchers who may plan to use a similar X-ray source and experimental setup.

\section{Experimental Setup}
\label{sec:Exp_Setup}

 \begin{figure}
 \centering
   \begin{subfigure}
     \centering
   \includegraphics[trim=10 10 10 10,clip,width=.5\linewidth]{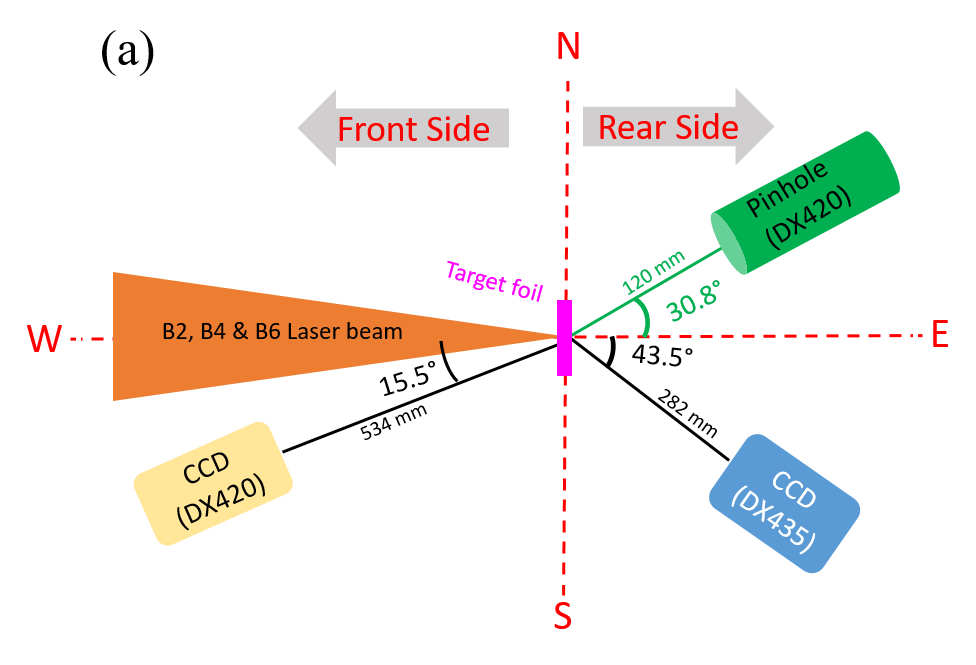}
   \end{subfigure}
   \hfill
  \begin{subfigure}
       \centering
     \includegraphics[trim=10 0 0 0,clip,width=.45\linewidth]{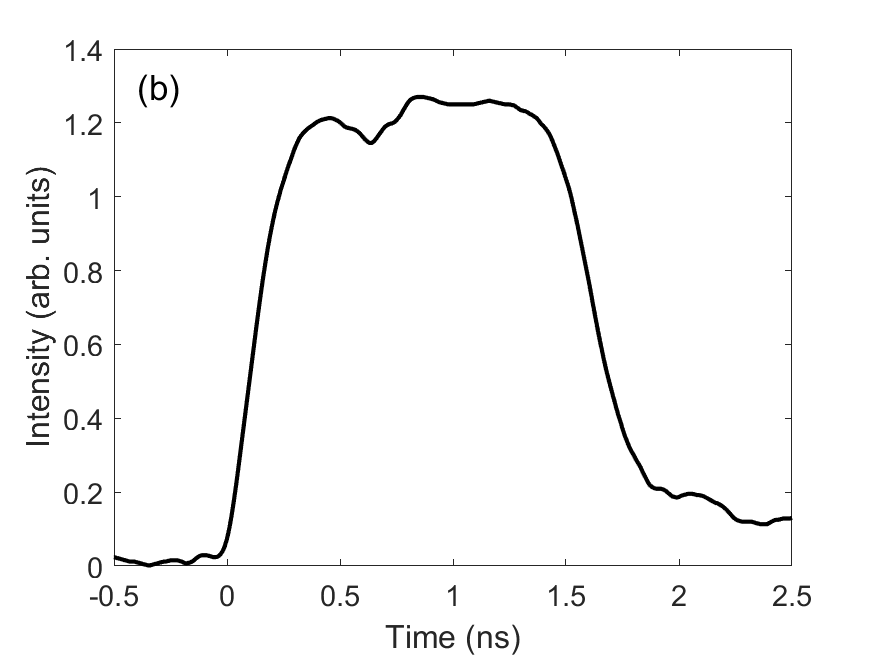}
  \end{subfigure}
  \caption{(a) A top-view schematic of the VULCAN experimental setup, with the Sn or Ag foil target placed in the vacuum chamber. A flat Si spectrometer coupled with an Andor DX435 detector is placed on the rear side of the target foil at 43.5$^{\circ}$. Another flat Si spectrometer coupled with an Andor DX420 detector is viewing from the front side of the target foil at an angle of 15.2$^{\circ}$. One pinhole setup coupled with an Andor DX435 detector is placed on the rear side at an angle of 30.8$^{\circ}$. All the angles are measured from normal to the target plane. An X-ray streak camera (XRSC) recorded both spectral and temporal emission (not shown in the figure). (b) A typical pulse shape for the incident laser beams recorded with an optical streak camera.}
  \label{fig:Experimental_Setup}
\end{figure}

A schematic of the VULCAN experimental setup is shown in figure  \ref{fig:Experimental_Setup}a. Three overlapping, frequency doubled beams were focused on the target (Sn or Ag) foil to deliver up to 500 J of laser energy at 527 nm, hence generating a multi-keV X-ray source. We took a number of data shots for both Sn (\uppercase{z}=50) and Ag (\uppercase{z}=47) target foils, varying the thickness of the foil and the laser spot size. These foils (Sn: 251/538/802 nm; Ag: 467 nm) were coated onto a CH layer of thickness 18.6 $\mu$m. The laser pulse Full-Width at Half-Maximum (FWHM) duration was approximately 1.5 ns with peak intensities varying from $~0.1-4 \times 10^{15}$ Wcm$^{-2}$. A typical pulse shape of the laser is shown in figure \ref{fig:Experimental_Setup}b. The laser intensity was mostly changed by varying the laser spot diameter, from 100 to 500 $\mu$m.

We recorded the Ag L-shell emission with two flat Si (111) crystal spectrometers coupled with Andor X-ray CCD detectors. The Si crystals were calibrated previously using an X-ray K-$\alpha$ source, with a resulting $\pm$10$\%$ uncertainty in the reflectivity \cite{Brozas2018}. One Si spectrometer was placed to view the front, laser-irradiated, side of the target at 15.2$^{\circ}$ to the target while the second was employed to view the rear side at an angle 43.5$^{\circ}$ to normal. Spectral ranges for this setup were $3670-4360$ eV for Sn foil shots and $3220-3680$ eV for the Ag foils. However, the available geometry within the chamber meant that the Si spectrometer on the front side was set to record a narrower range, namely $3740-4020$ eV for Sn and $3270-3500$ eV for Ag.
\begin{figure}
	\centering
	\includegraphics[trim=10 0 10 10,clip,width=0.7\linewidth]{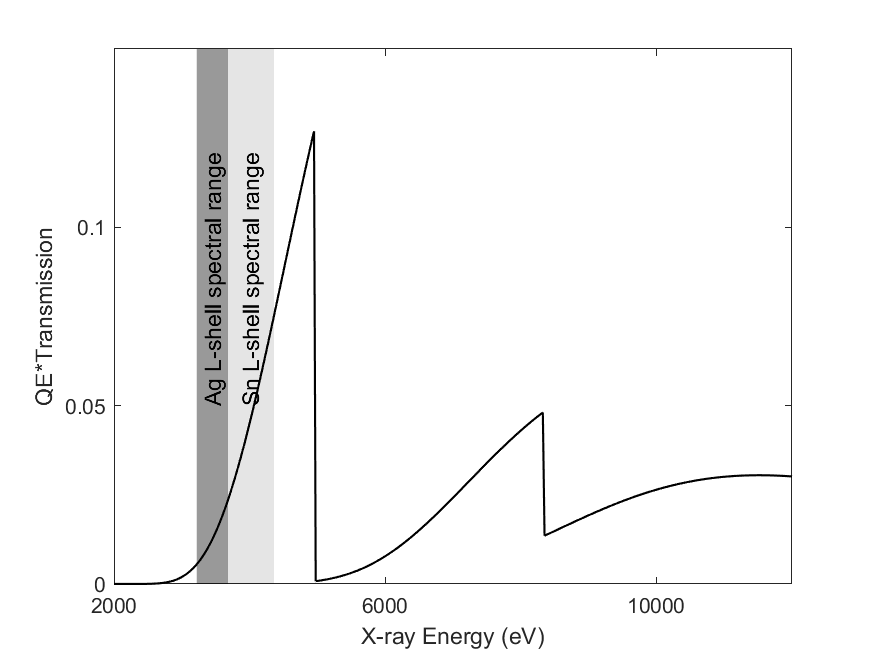}
  \caption{Total response for the pinhole including the CCD response, filtering and CH on the target. The Ag and Sn L-shell spectral ranges recorded on the rear spectrometer are indicated as dark and light grey shadings on the plots.}
  \label{fig:pinholefilter}
\end{figure}
The pinhole camera was used with a 10 $\mu$m diameter pinhole in a tantalum substrate and was connected to an Andor DX435 CCD. It was set up to view the rear side at 30.8$^{\circ}$ from normal in order to record the spatial distribution of the generated X-ray source. The magnification of the pinhole image was 3.92 $\pm$ 0.03. A combination of 5 $\pm$ 1 $\mu$m Ni and 17.80 $\pm$ 0.02 $\mu$m Ti foil filters was used, which cut out softer, sub-keV, X-rays but also has K-edges to cut transmission at photon energies above the L-shell. We have determined the relative spectral response of the pinhole by combining the transmission of the filters and  quantum efficiency of the CCD and present this in figure \ref{fig:pinholefilter}. As we can see, the peak response is above the L-shell photon energies, at around 5 keV, with some response out to beyond 10 keV. Nevertheless, there is a reasonable relative sensitivity in the L-shell photon range and the focal spot sizes measured should give a sufficiently accurate estimate of the L-shell source size.

Finally, an X-ray streak camera (XRSC) coupled to a HOPG crystal viewed the rear of the foil at an angle of 20$^\circ$ from normal to measure the X-ray temporal profile for a limited spectral range of L-shell emission. The streak speed and slit width (1 mm) were chosen to give roughly 100 ps temporal resolution. 

\section{Analysis and Results}
\label{sec:Data_Ana}

\subsection{L-shell yield}
\begin{figure}
 \centering
   \begin{subfigure}
     \centering
   \includegraphics[width=.8\linewidth]{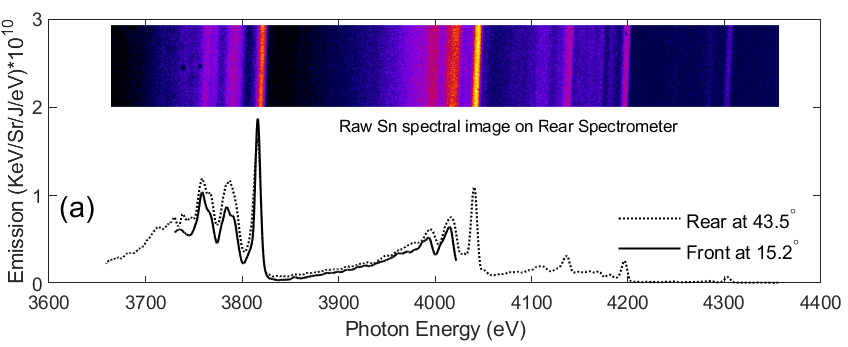}
   \end{subfigure}
   \hfill
  \begin{subfigure}
       \centering
     \includegraphics[width=.8\linewidth]{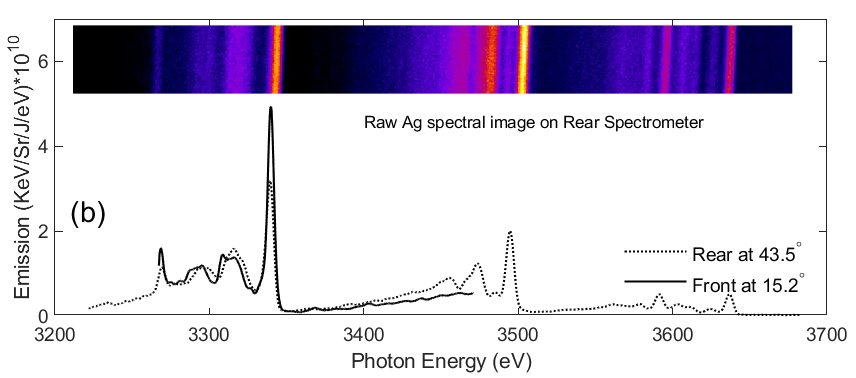}
  \end{subfigure}
  \caption{(a) Line-outs of Sn L-shell emission recorded on the front and rear Si spectrometers. A typical Sn spectral raw image from the rear spectrometer is shown in the inset. Thickness of the Sn foil for this shot was 251 nm. (b) Line-outs of Ag L-shell emission recorded on front and rear Si spectrometers. A typical Ag spectral raw image from the rear spectrometer is shown in the inset. All the spectral profiles are corrected for filtering and the quantum efficiencies of the CCDs.}
  \label{fig:Spect_profile}
\end{figure}

Typical spectral raw images and line-outs, corrected for filtering, CCD quantum efficiency and crystal response, for Sn and Ag foils are presented in figures \ref{fig:Spect_profile}a and \ref{fig:Spect_profile}b, respectively. What is not evident in these figures is that an X-ray block was present on the spectrometer housing, to mask a strip of $\sim$ 50 pixels for each of the CCD cameras. This provided a region where the background due to crystal/filter fluorescence and hard X-rays not diffracted from the crystal would be present, and hence allow a background subtraction from the data. The L-shell Sn spectrum from the experiment has been spectrally calibrated by the use of spectra from previous work in White et al \cite{White2018}.
We calibrated the L-shell Ag spectra using a calibration shot with KBr, in which the He-$\alpha$ line group (1s$^{2}$-1s2p $^{1}$P and $^{3}$P lines and Li-like satellites) was recorded. This yielded a spectral calibration that matched the expected dispersion within a fraction of a percent.
 
We corrected the obtained spectral data for signal background, the energy-dependent transmission of filters and quantum efficiency of the CCD, the latter taken from the manufacturer's data.  The effective solid angle for collection, $d\Omega_{eff}$ for a flat crystal spectrometer is given by;

\begin{equation}
\label{eqn:solidangle}
d\Omega_{eff} = R_{c}\frac{W{ccd}}{L}   
\end{equation}

where $R_{c}$ is the integrated reflectivity for the crystal \cite{henke1993}, which only varies slightly with energy over the range of interest. As noted above, the integrated reflectivities were confirmed to $\pm$ 10$\%$ experimentally using an  X-ray source \cite{Brozas2018}. $W_{ccd}$ is the width of the CCD chip and $L$ is the total path distance from the source to the detector via the crystal. The counts per pixel can be converted to photons using the A/D conversion of the CCD, which was also determined from \cite{Brozas2018} and the energy per electron-hole pair of 3.65 $\pm$ 0.03 eV, which is known e.g. \cite{kraft1995}.  This allows us to convert the total counts on the CCD to energy per steradian and with the known dispersion we obtain the energy per steradian per Joule of laser energy, per unit photon energy as displayed in figure \ref{fig:Spect_profile}. 

 \begin{figure}
 \centering
   \begin{subfigure}
     \centering
   \includegraphics[trim=15 0 10 0,clip,width=.4\linewidth]{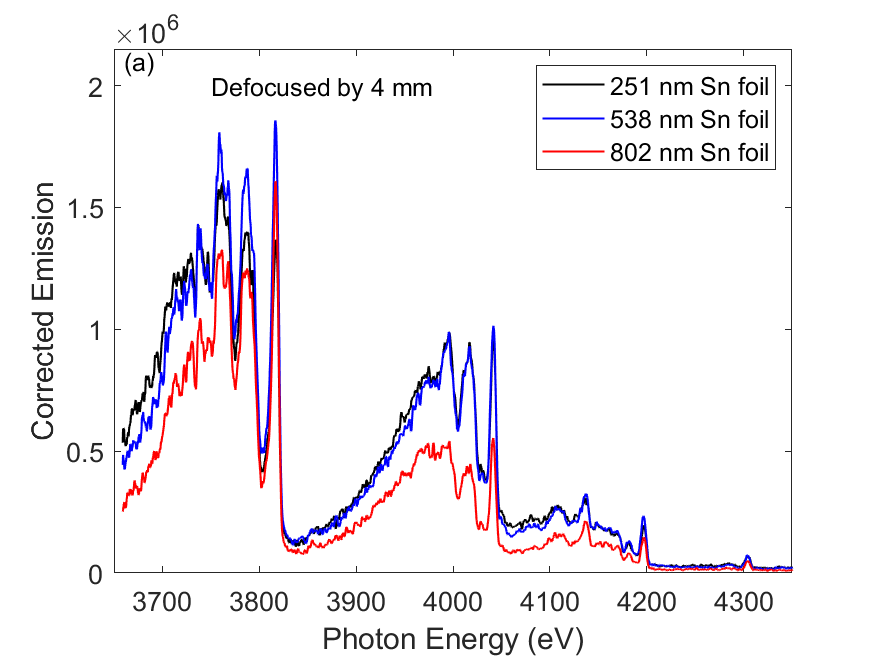}
   \end{subfigure}
    \begin{subfigure}
       \centering
     \includegraphics[trim=15 0 10 0,clip,width=.4\linewidth]{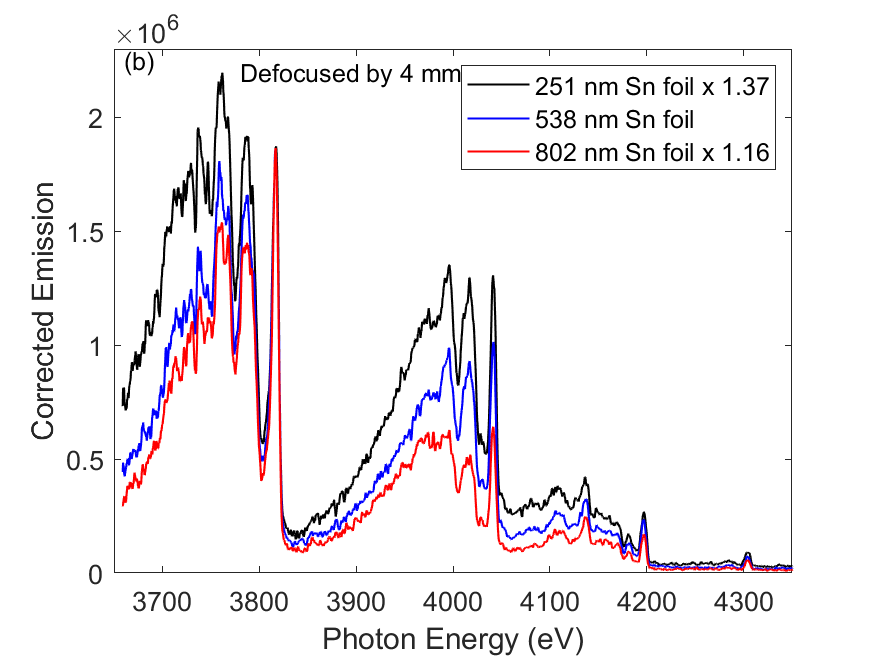}
  \end{subfigure}
    \begin{subfigure}
       \centering
     \includegraphics[trim=15 0 10 0,clip,width=.4\linewidth]{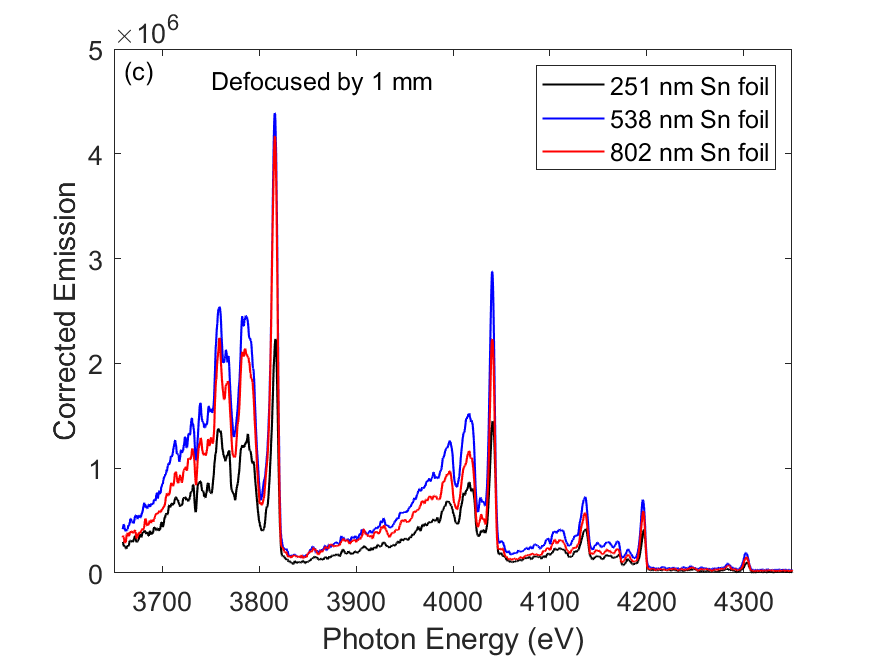}
  \end{subfigure}
  \begin{subfigure}
       \centering
     \includegraphics[trim=15 0 10 10,clip,width=.4\linewidth]{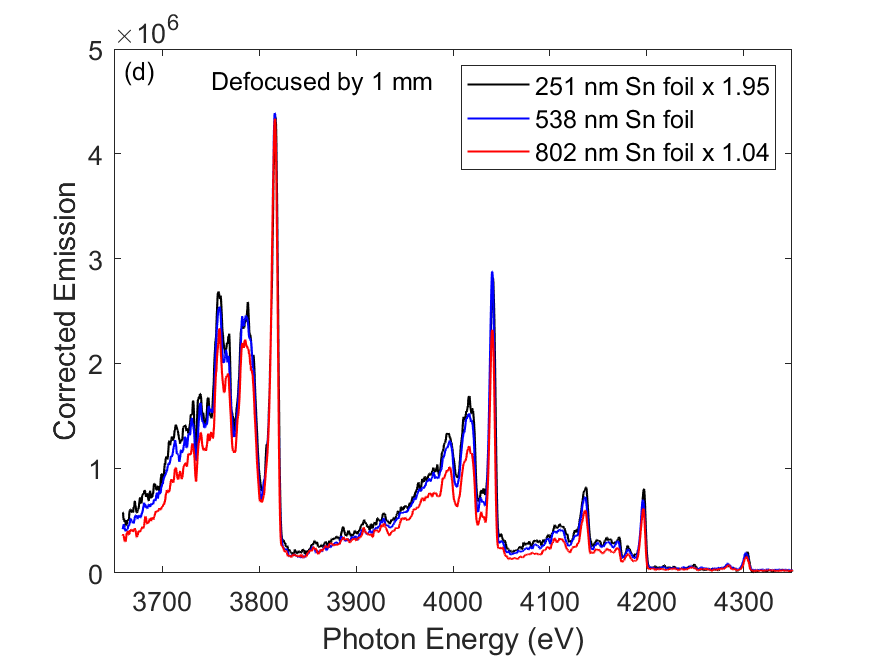}
  \end{subfigure}
  \caption {(a): Experimental sample L-shell spectra for Sn of thickness 251, 538 and 802 nm at lens’s defocused position 4 mm. (b): Scaled Sn 251 nm and 802 nm thickness spectra with respect to 538 nm presented in (a). (c): Experimental sample L-shell spectra for Sn of thickness 251, 538 and 802 nm at lens’s defocused position 1 mm. (d): Scaled Sn 251 nm and 802 nm thickness spectra with respect to 538 nm presented in (c).}
  \label{fig:Sn_Spect_profile_thick}
\end{figure}

The dependence of the spectral shape on the Sn thickness can be seen in figure \ref{fig:Sn_Spect_profile_thick}. We performed several shots to record Sn spectra at different foil thickness (251 nm, 538 nm and 802 nm) by varying the defocus position of lenses from 1-4 mm. The mm defocus distances refer to defocus of the f/10 lens so we are going from a 100 $\mu$m to a 400 $\mu$m focal spot. In figure \ref{fig:Sn_Spect_profile_thick} we present the two extreme defocus positions. In the left-hand panels (a) and (c) we plot the profiles corrected for filters etc, recorded on the rear spectrometer for thicknesses 251 nm, 538 nm and 802 nm, while in (b) and (d) we scale the 251 nm and 802 nm Sn emission profiles to match the strong line at 3816 eV with 538 nm Sn thickness. We include the multiplication factor needed for the spectra to be normalised to this feature. Features at lower energies than the 3816 eV line are reduced in intensity with increasing thickness, as are the whole line-groups at higher energy. We do notice that the effect is reduced as we move to higher intensity (smaller focal spot). The change is spectrum is presumably due to variations in opacity with thickness. We expect that there will be a strong temperature gradient in the foil. Looking in the rear direction, for much of the pulse, there will be a region slightly cooler than the emitting layer but containing a high population of the emitting ions, in their ground state. This absorbing layer will increase with foil thickness. For a smaller focal spot, we expect higher temperature and a faster burn-though of the foil and this would limit the effect of a cooler absorbing layer.

\begin{figure}
  \centering
    \includegraphics[width=0.8\linewidth]{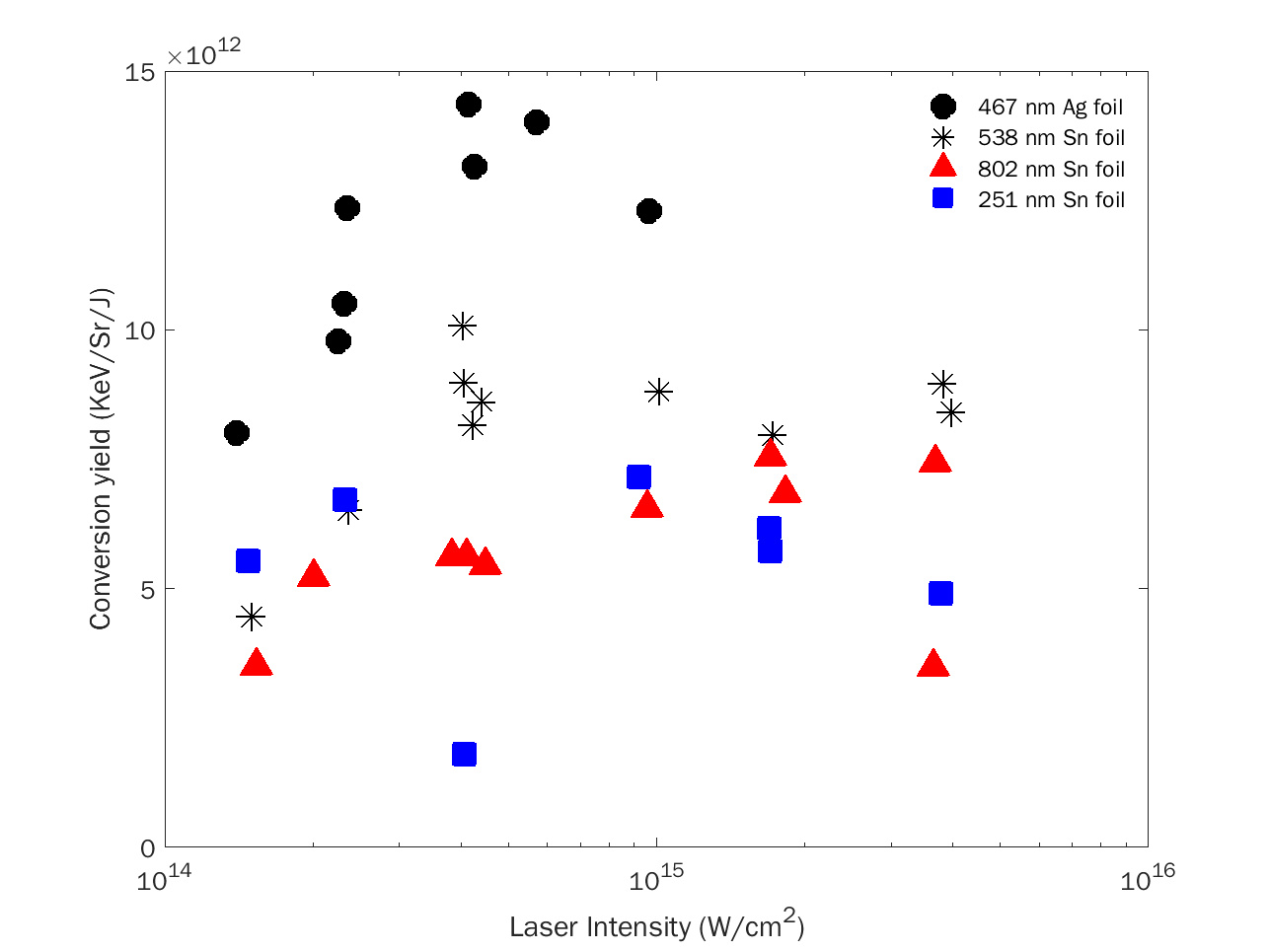}
  \caption{Laser energy (J) to X-ray energy (keV) conversion yield measured for Sn and Ag L-shell emission recorded on the rear side (at $43.5^{\circ}$) Si spectrometer.}
  \label{fig:efficiency_rear}
\end{figure}

In figure \ref{fig:efficiency_rear} we plot the spectrally integrated laser to X-ray conversion yield (keV/Sr/J) against input laser intensity, for the L-shell, recorded on the rear side Si spectrometer for both Ag and Sn targets. As we expected, the conversion yield for the Ag targets is higher than that for Sn.  We can see that, for Sn, the conversion efficiency drops for both the lowest and highest thickness of foil. This might indicate that burn-through is completed early for the 251 nm foils, and this interpretation is supported by the shorter duration of the X-ray pulse seen below. For the 802 nm foils, there is no increase in pulse duration but the yield is a little smaller. This may indicate that, at the optimum intensity, the heatwave does not burn through significantly more than for the 538 nm foil and that the lower yield may be due to the L-shell emission being partially absorbed by a lower temperature layer, which is not sufficiently hot to emit L-shell radiation but has some opacity to those X-rays. Thus, the optimum thickness for yield may be broadly similar to the 538 nm foil thickness. Similar results are presented in Wang et al \cite{WANG2021} for similar Sn thicknesses. There is some variability in yield, as is common for such experiments, but the optimum intensity in each case appears to be $\sim$ 4$\times10^{14}$ Wcm$^{-2}$.

\begin{figure}
  \centering
   \includegraphics[width=0.8\linewidth]{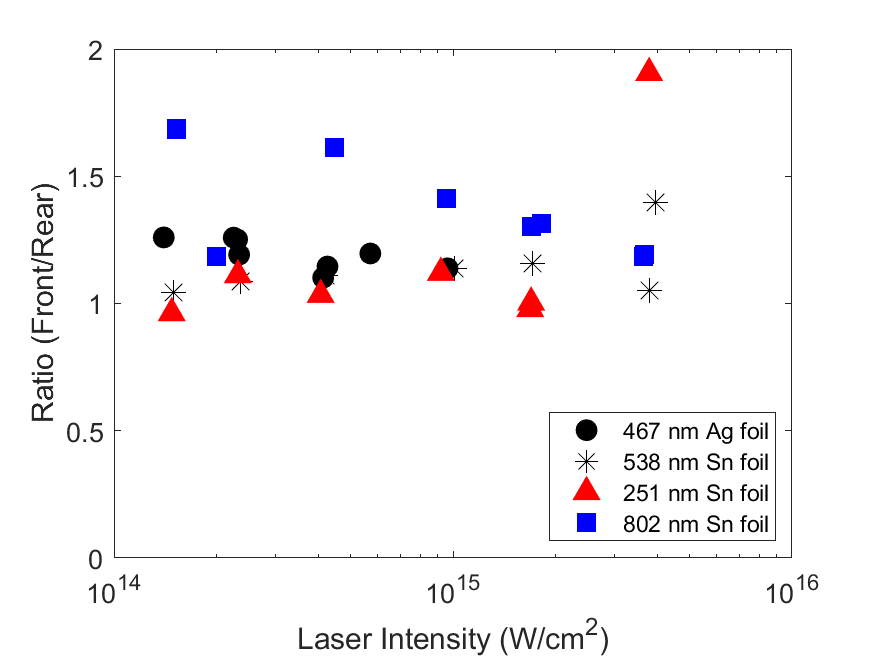}
  \caption{ Ratios of experimental conversion yields for the front and rear spectrometers at 15.2 and 43.5 ${^\circ}$, respectively. Given the $\pm$10$\%$ systematic error bars on the calibration of the Si crystals, the ratio is approximately unity as predicted in the simulation below.}
  \label{fig:frontrearratio}
\end{figure}

As noted above, the front Si spectrometer at 15.2$^{\circ}$ from normal has a comparatively narrow spectral range. In figure \ref{fig:frontrearratio}  we compare the conversion yields for the front and rear spectrometers for Ag and Sn shots, determined for the common spectral features recorded by both spectrometers. For Sn the spectral range is approximately $3740-4020$ eV, and $3270-3500$ eV for Ag. As we can see, the ratio is close to unity for much of the data, which is slightly counter-intuitive, given the rear spectrometer is further from the normal to the target, but as discussed below, is in broad agreement with simulation. Again, there is some scatter in the data, however, one can discern a tend downwards for the 802 nm foil as intensity increases. This might be expected if the burn-through is not complete because there would be a colder non-emitting but absorbing layer ahead of the rear-side emission and this would narrow as the intensity and thus burn-through depth increased. This interpretation is consistent with the observation that, for the 251 nm  Sn foils, barring one outlier, the ratio is consistently lower than for the thickest foils.

\subsection{X-ray streak data}
\begin{figure}
  \centering
   \includegraphics[trim=0 145 10 0,clip,width=1\linewidth]{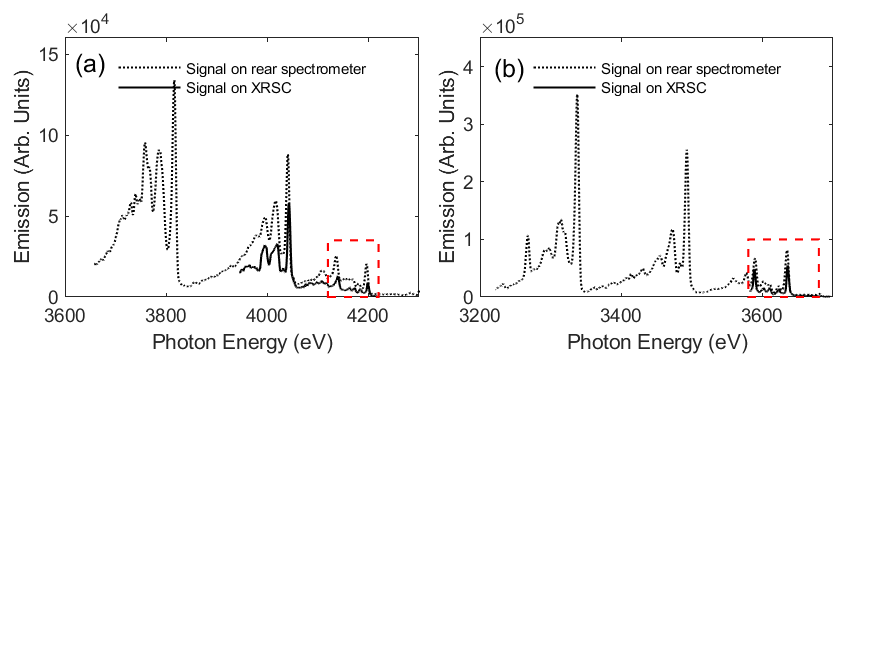}
  \caption{Comparison of typical L-shell spectral profile recorded on XRSC and Si (rear side) spectrometer setups for (a) Sn and (b) Ag target foils. The common spectral features are highlighted with a red rectangle. We use arbitrary units as the streak data is not absolutely calibrated.}
  \label{fig:XRSCspectralprofile}
\end{figure}

We identified the spectral features in the XRSC data by comparison with the spectrum from the flat spectrometer on the rear side at 43.5$^{\circ}$. We did not have an absolute calibration for the streak camera efficiency and so do not obtain yield data but do have the temporal history. The distance of the streak camera from the target means that the L-shell spectral features recorded on the XRSC setup do not cover the complete spectrum. In figure \ref{fig:XRSCspectralprofile} we show spectral line-outs for both Sn and Ag foils, where we compare the spectral windows of the streak data with those obtained from the flat-crystal instruments. Our streak camera measurements are for a selected part of the L-shell spectrum, but the relatively narrow total range of the L-shell spectrum means we can take the results as typical as far as temporal history is concerned.

\begin{figure}
\begin{subfigure}
  \centering
  \includegraphics[trim=0 0 30 10,clip,width=0.45\linewidth]{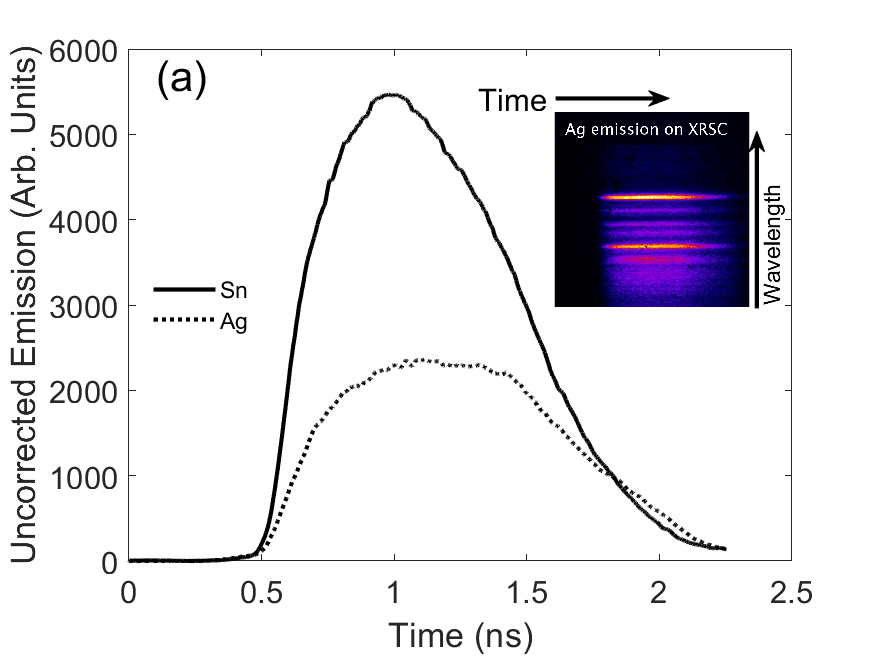}
  \label{fig:sub1}
\end{subfigure}
\begin{subfigure}
  \centering
  \includegraphics[trim=10 0 10 10,clip,width=0.45\linewidth]{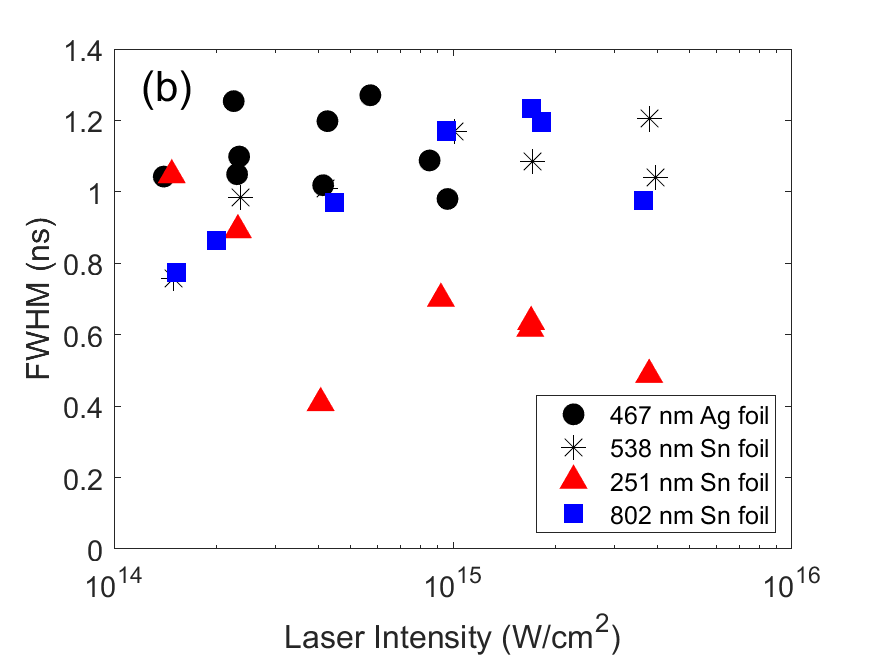}
  \label{fig:xrscdata}
\end{subfigure}
\caption{(a) Typical raw XRSC temporal profiles recorded for the Ag and Sn target foils. A typical raw Ag XRSC image is shown in the inset. (b) FWHM of XRSC temporal profile for each shot plotted against laser intensity.}
\label{fig:test}
\end{figure}

In figure \ref{fig:test}(a), we present temporal L-shell histories for both the Ag and Sn foils, indicating typical FWHM of $\sim$1.1 ns in both cases. The X-ray streak is triggered by a signal coming from the laser and so is synchronised to the laser. However, in timing the camera, we adjust to capture the X-ray emission, which we expect to begin after the start of the laser pulse, by a time of the order of the laser rise-time. Despite some shot-to-shot variation, the FWHM of the emission has a weak dependence on laser intensity for the thicker foils, as shown in figure \ref{fig:test}(b). We can also see that the results for the 538 nm and 802 nm Sn foils are very similar even at the highest irradiances used. On the other hand, the 251nm foils show a consistently decreasing  duration as intensity is increased. This indicates early burn-through with the Sn foil ablating to well below the critical density and cooling before the end of the laser pulse. For the other foils, there is a general increase in duration with intensity, but the lack of a clear increase for the 802 nm foils compared to the 538 nm foils, indicates still that burn-through is not extending significantly beyond the latter.

\subsection{Pinhole data} 
\begin{figure}
  		\centering
     \includegraphics[trim=0 130 70 0,clip,width=0.7\linewidth]{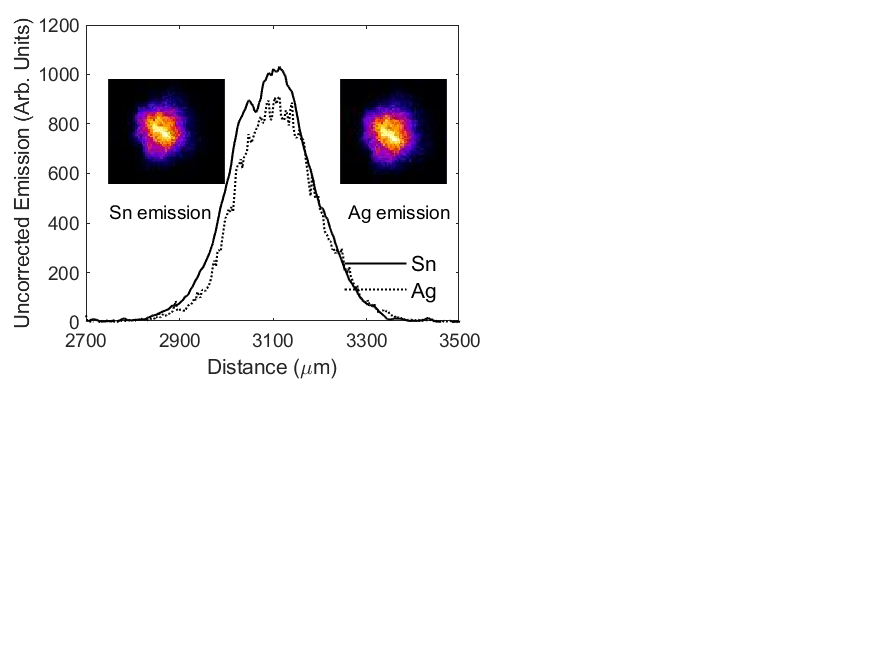}
  \caption{Typical raw image recorded on the pinhole, and its line-outs for the Sn and Ag foils. The nominal FWHM diameter of the laser spot in both cases is $\sim$ 200 $\mu$m since the focal offset was 2 mm for f/10 lenses.}
  \label{fig:pinhole_lineouts}
\end{figure}

We analysed the pinhole data recorded for each shot for the Sn and Ag target foils and corrected the data for the background signal surrounding each image. Typical pinhole raw images and their line-outs for Sn and Ag foils are shown in figure \ref{fig:pinhole_lineouts}. We determined the object size from the magnification of the pinhole setup, as previously noted. For both the cases presented, the laser focusing was offset from optimal by 2 mm, thus there should be a nominal 200 $\mu$m optical focal spot. The X-ray images of L-shell emission in figure \ref{fig:pinhole_lineouts} have a FWHM similar to this for both cases, with $\sim$190 $\mu$m for Ag and $\sim$200 $\mu$m for Sn.

In principle we should be able to estimate the yield from the pinhole images. This was attempted by considering the filter transmission, the CCD response and measured L-shell spectrum. However, the estimates fell about a factor of three lower than those derived from the crystal spectrometer data and significantly below the values from simulations. This is possibly due to a combination of uncertainties in the filter thicknesses, pinhole size and a limited measurement range for the incident spectrum. 

\section{Plasma Simulations and Discussion of Results}
\label{sec:Result}

\begin{figure}
  \centering
   \includegraphics[trim=0 0 0 0,clip,width=0.6\linewidth]{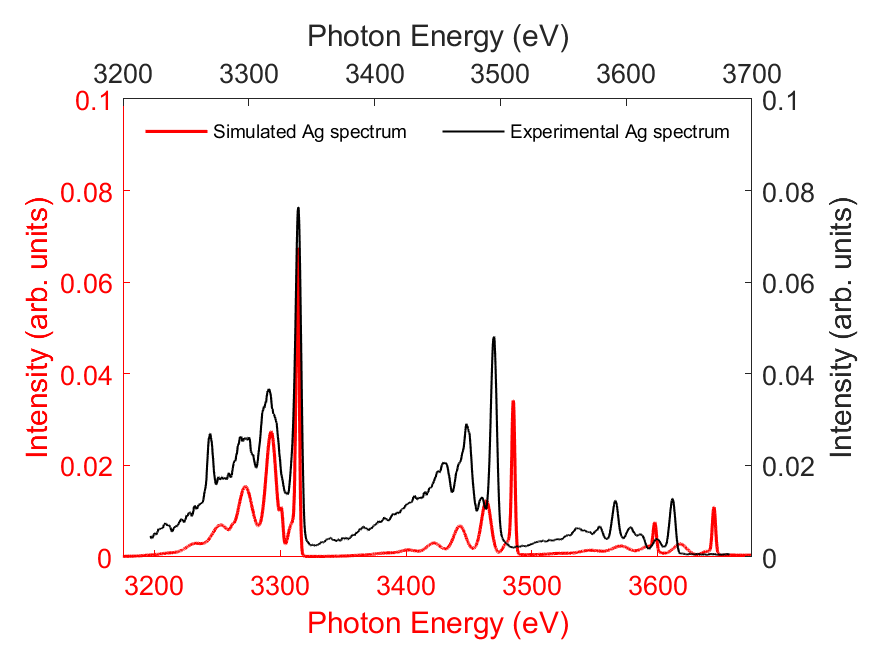}
  \caption{Comparison of sample experimental Ag spectrum with a FLYCHK simulation, generated at an electron temperature 2000 eV and electron density 10$^{21}$cm$^{-3}$.}
  \label{fig:sim_spectra}
\end{figure}

In figure \ref{fig:sim_spectra}  we show an experimental Ag spectrum compared to a simulated spectrum created with the FLYCHK code \cite{Chung2005}, generated assuming an electron temperature of 2000 eV and electron density of 10$^{21}$ cm$^{-3}$. As we see, there is a spectral shift between our data and FLYCHK. This arises due to the fact that FLYCHK does not use a detailed data base for L-shell transitions but rather a Moseley type scaling law, which is accurate to better than 1\% (about 25 eV in this case). However, the features are broadly as expected. The experimental spectra presented here closely match those for Ag presented in Hu et al \cite {Hu2008}. Simulated spectra for Sn foils may be found in Bailie et al \cite{BAILIE_2020}. 

We can compare our experimental results with a simulation that we have undertaken with the 2D NYM radiation-hydrodynamics code \cite{Roberts1980}. The atomic physics was calculated in each Lagrangian cell and coupled self-consistently to the rad-hydro using a non-LTE model which was based on the XSNQ code \cite{Grasberger1965,Grasberger1966,LokkeandGrasberger1977}. However, this was updated to include auto-ionisation and di-electronic recombination by Rose et al. \cite{rose2004} using a model based on Albritton and Wilson \cite{AlbrittonandWilson1999,AlbrittonandWilson2000}. The radiation transport employed was multi-group implicit Monte-Carlo. All NYM results are Crown Copyright.

\begin{figure}
  \centering
   \includegraphics[trim=0 145 10 0,clip,width=1\linewidth]{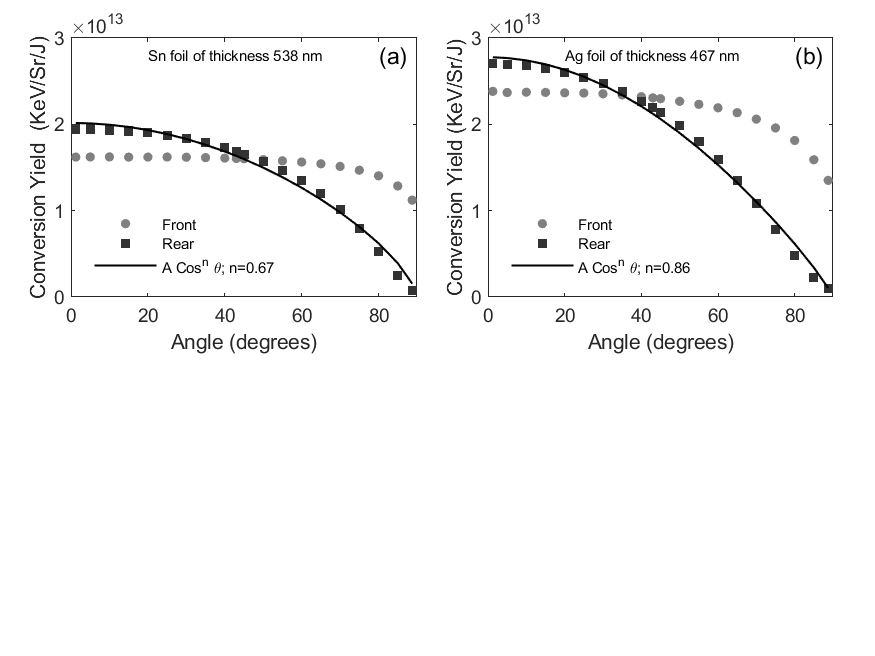}
  \caption{(a) Front and rear simulated conversion yield with respect to angle, cosine fit included, for Sn 538 nm target foil; (b) same as (a) but for Ag 467 nm target foil.}
  \label{fig:sim_angular}
\end{figure}

We investigated the angular dependence of X-ray emission coming from the target foil through simulation within the emission energy range of 3.2-4.4 keV at laser intensity $4.8\times{10}^{14}$ Wcm$^{-2}$. This was chosen as it is close to the optimum intensity seen in the experimental data. We used the same target foils as in our experiments, Sn (of thicknesses 251 nm, 538 nm and 802 nm) and Ag (of thickness 467 nm). Simulation results for the 538 nm Sn and 467 nm Ag foils are shown in figure \ref{fig:sim_angular} for both the front and rear emission. We see that, similar to the experimental results, the conversion yield for Ag is higher than for Sn. The angular variation of X-ray emission is also quite isotropic on the front side, whereas it has a broadly cosine dependence on the rear side of the target for each case.  We have fitted the rear-side angular conversion yield with functions of the form $A$ cos$^n\theta$, where n=0.67 and 0.86 for the Sn and Ag foil data, respectively. For the rear emission at 43.5$^{\circ}$, the simulated Sn emission is 1.6$\times$10$^{13}$ keV/Sr/J, which is about 70$\%$ higher than seen experimentally. In the case of Ag, the rear yield is 2.2$\times$10$^{13}$ keV/Sr/J, which is approximately 60$\%$ higher than experiment. The higher yield in the simulations may be a result of the fact that the experimental laser focal spot is not smooth, as in the simulation. There will be non-uniformities or 'hot spots' which means that, even if the average intensity is optimal, much of the energy will be incident at either higher or lower intensities. 

Averaging over all angles, our simulations indicate the L-shell emission from the rear to be 1.3$\%$ of the laser energy for the 538 nm Sn foil, compared to 2.0$\%$ for the front side. For the Ag foil simulations, the equivalent yields are 1.6$\%$  and 2.7$\%$ respectively. If we make the assumption that the experimental angular variation is similar, then scaling the comparison at angles for which we have measurements, indicates an overall conversion to the rear side of $\sim$0.7$\%$ for the Sn case and $\sim$1.0$\%$ for Ag.  We note that a previous study of conversion efficiency to Ag L-shell X-rays, using a nanosecond laser, by Hu et al \cite{Hu2008} reported 1.2-1.4\% conversion efficiencies for Ag targets on the front side of the target but with a thicker Ag foil (2 $\mu$m).

Interestingly, although the overall emission from the front side is higher, at small angles from the target normal, the front emission is predicted to be lower than the rear, with emission at the angles of our front and rear spectrometers predicted to be similar. This is, in fact, close to what we observe experimentally, as can be seen from figure \ref{fig:frontrearratio}. The predicted higher emission for the rear side at small angles from normal is not an intuitively obvious result. A broad explanation for this phenomenon may be that the expansion of the coronal plasma on the front (laser irradiated) side is expected to result in a plasma with a density scale-length in the region 100-200 $\mu$m. This means that, for our focal spot size, we do not expect a planar expansion. This may leave the possibility open that whilst the opacity for the front emission is higher than for the rear at normal incidence, thus explaining the result, the path length is varying far less with angle than for the rear direction. However, a full investigation would merit further experiments with a more detailed set of angular measurements.

\begin{figure}
  \centering
   \includegraphics[trim=0 145 10 0,clip,width=1\linewidth]{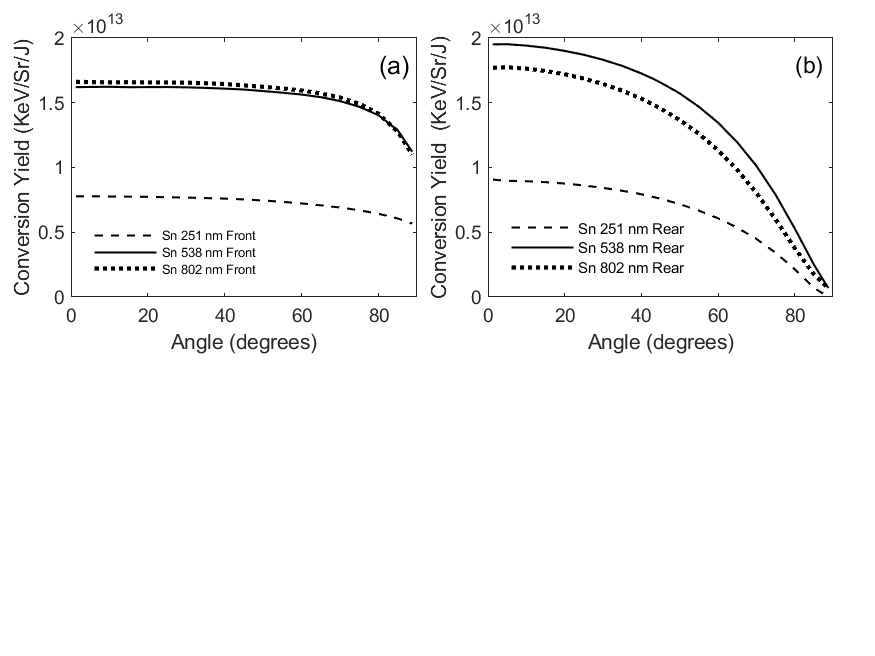}
  \caption{(a) Front side NYM code simulated conversion yield with respect to angle for different thicknesses of Sn target foil; (b) Same as (a) but for the rear side conversion yield.} 
  \label{fig:sim_Sn_thick}
\end{figure}

Figure \ref{fig:sim_Sn_thick} shows the variation of simulated conversion yield for different thicknesses of Sn foil. The front side emission barely increases from 538 nm to 802 nm, indicating that the former may, as discussed above, be close to optimum. This is supported by the rear side simulated emission which, as for the experimental data, drops a little for the 802 nm case, indicating a colder, more optically thick layer is present to the rear side, as discussed earlier. The yield for the 251 nm case has dropped by just over a factor of 2 compared to the 538 nm optimum, which is broadly consistent with the experimental observation seen in our data. Simulated data from NYM, presented in figure \ref{fig:sim_Sn_thick}b, are also broadly supportive of the experimental results reported in Bailie et al \cite{BAILIE_2020} for Sn targets, using 0.35 $\mu$m laser irradiation, where 0.9$\%$ overall conversion yield was seen.

\section{Conclusions}
\label{sec:Conlcusion}
In conclusion, our experimental results have demonstrated that Ag produces a higher conversion yield overall than Sn. This applies to both the front and rear sides of the targets with respect to the illuminating laser, a useful result for planning experiments where the X-ray emission is being used to photoionise or heat a gas or foil target. The simulations produced comparable results, within a factor of 2, to our experimentally determined conversion yields from our front and rear Si spectrometers, with 538 nm Sn being an optimum (out of the three) thicknesses. Experimental results show that the conversion yield rises with intensity but plateaus at around 10$^{15}$ Wcm$^{-2}$, suggesting this intensity is optimal for both Ag and Sn shots. Simulations indicate that the angular variation of emission behaves differently on the front and rear sides of the target. It is mostly isotropic on the front side whereas broadly cosine dependent on the rear. The results of the simulation indicate that more detailed measurements of the angular variation should yield an interesting result regarding the ratio of front to rear emission at angles close to normal.

\section{Conflicts of Interest}
The authors declare that there are no conflicts of interest regarding the publication of this article.

\section{Acknowledgements} 
This work was supported by the UK Science and Technology Facilities Council grant ST/P000312/1. We would like to thank the Central Laser Facility staff running the laser, target area and target preparation facilities for their contributions.
\medskip

\bibliographystyle{unsrt}
\bibliography{paper}

\end{document}